\documentclass[fleqn,usenatbib,useAMS]{mnras}

\usepackage{graphicx}	
\usepackage{amsmath}	
\usepackage{amssymb}	
\usepackage{multicol}   
\usepackage[dvipsnames]{xcolor}
\usepackage[british]{babel}
\usepackage{csquotes}
\usepackage{changepage}
\usepackage{caption}
\usepackage[export]{adjustbox}
\usepackage{mwe}
\usepackage{float}
\usepackage{multirow}
\usepackage{bold-extra}
\usepackage{bm}	 

\usepackage[T1]{fontenc}
\usepackage{ae,aecompl}

\title[Time lag properties of HFQPOs in GRS 1915+105]{First detection of Soft-lag in GRS 1915+105 at HFQPO using \textnormal{\textit{AstroSat}} observations}

\author[Majumder et al.]{Prajjwal Majumder$^{1}$
\thanks{E-mail: \href{mailto:majumderprajjwal@gmail.com}{majumderprajjwal@gmail.com}}
, Broja G. Dutta$^{1}$
\thanks{Contact E-mail: \href{mailto:brojadutta@gmail.com}{brojadutta@gmail.com}}%
, Anuj Nandi$^{2}$\thanks{E-mail: \href{mailto:anuj@ursc@gov.in}{anuj@ursc.gov.in}}
\\
$^{1}$Rishi Bankim Chandra College, Naihati, West Bengal,743165, India \\
$^{2}$Space Astronomy Group, ISITE Campus, U. R. Rao Satellite Centre, Outer Ring Road, Marathahalli, Bangalore, 560037, India.
}

\date{Accepted XXX. Received YYY; in original form ZZZ}

\pubyear{2023}

\begin{document}
\label{firstpage}
\pagerange{\pageref{firstpage}--\pageref{lastpage}}
\maketitle

\begin{abstract}
The Galactic black hole GRS 1915+105 exhibits generic High-Frequency Quasi-periodic Oscillations (HFQPOs) at $\sim$ 67 Hz only during the radio-quiet `softer' variability classes.
We present the time-lag properties associated with HFQPOs in the wide energy band (3$-$60 keV) using all \textit{AstroSat} observations.
For the first time, we detect soft-lag of 6$-$25 keV band w.r.t 3$-$6 keV band for all `softer' variability classes ($\delta$, $\omega$, $\kappa$ and $\gamma$). Moreover, our findings reveal that soft-lag increases gradually with the energy of the photons. These features are entirely opposite to the previous report of hard-lag obtained with the \textit{RXTE} observations.
The energy-dependent time-lag study exhibits a maximum soft-lag of $\sim$ 3 ms and $\sim$ 2.5 ms for the $\delta$ and $\omega$ classes respectively, whereas the $\kappa$ and $\gamma$ classes both exhibit a maximum soft-lag of $\sim$ 2.1 ms.
We find a coherent lag-energy correlation for all four variability classes, where the amplitude of soft-lag increases with energy and becomes maximum at $\sim$ 18 keV. We interpret this observed soft-lag as the reflection of hard photons in the `cooler' accretion disc. A generic lag-rms correlation implies that the soft-lag increases with the rms amplitude of the HFQPO. The wideband (0.7$-$50 keV) spectral study suggests a high value of the optical depth ($\tau$ $\sim$ 6.90$-$12.55) of the Comptonized medium and the magnitude of the soft-lag increases linearly with the increase in optical depth ($\tau$). We explain the observed time-lag features at the HFQPOs in the context of a possible accretion disc scenario.

\end{abstract}




\begingroup
\let\clearpage\relax
\endgroup
\newpage

\section{Introduction} \label{introduction}

X-ray binaries (XRBs) often exhibit excess power in certain frequencies called Quasi-periodic Oscillations (QPOs) in its power density spectrum (PDS) \citep{van_der_Klis1985}.
In general, the QPOs in black hole XRBs (BH-XRBs) appear in the range of 0.1$-$450 Hz. QPOs can be classified depending on the frequency range in two general categories, Low-frequency QPO (LFQPO) and High-frequency QPO (HFQPO) with centroid frequency < 40 Hz and 40$-$450 Hz respectively \citep{remillard_mcClintock2006}. LFQPOs are observed commonly in many BH-XRB systems, whereas HFQPOs are detected in a few sources. HFQPOs are important as their frequency is in the range expected for its Keplerian motion near the black hole \citep{stella_vietri1998,merloni1999,rebusco2008,stefanov2014} and it might also be generated due to the oscillations of the compact corona \citep[and references therein]{seshadri2022}. 
HFQPOs were observed in a few sources such as GRS 1915+105 \citep{morgan1997,belloni2013,belloni2019,sreehari2020,seshadri2022}, GRO J1655-40 \citep{remillard1999,strohmayer2001a}, XTE J1550-564 \citep{miller2001}, H 1743-322 \citep{homan2005,remillard2006}, IGR J17091-3624 \citep{altamirano2012}. In some cases, simultaneous observations of HFQPOs in $\sim$ 3:2 frequency ratio were observed in GRO J1655-40, XTE J1550-564, GRS 1915+105 and H 1743-322. GRS 1915+105 has also shown a second pair of HFQPO which are not in a 3:2 ratio \citep[41 Hz and 67 Hz,][]{morgan1997,strohmayer2001b}. These HFQPOs are of further interest because their centroid frequency does not drift very much in response to considerable luminosity changes. This is an important difference between the HFQPOs of black hole binary and the variable kHz QPOs of neutron stars \citep{remillard_mcClintock2006}. Often the presence of prominent HFQPOs can be observed during the soft/soft-intermediate states of the source, dominated by the disc emission \citep{belloni2012}. 

The BH-XRBs exhibit complex non-linear time-lag features which can reveal the dynamics of accretion  geometry i.e., the variation of the size of the Comptonizing region. 
The energy-dependent variation of time-lag can explain the responsible physical mechanisms that produce HFQPO. Time-lag represents the time difference in arrival time between soft photons and hard photons. The delay in the arrival of hard photons as compared to soft photons is termed as hard-lag and vice-versa \citep{van_der_klis1988_lag}. Time-lag was initially thought to be generated due to the inverse Comptonization of soft seed photons by hot electrons  \citep{payne1980,miyamoto1988} which only explains the hard-lag. Several models have been proposed \citep{cui1997_cyg-x1,nowak1999,poutanen2001} to explain the hard and soft-lag observed in Galactic black hole candidates. \cite{reig2000} suggested that both hard-lag and soft-lag can be explained by Compton up-scattering and down-scattering mechanisms. 
\cite{qu2010} reported an anticorrelation between phase-lag and LFQPO frequency (0.5$-$10 Hz) for GRS 1915+105 where, the lag switches sign from positive to negative at the QPO frequency $\sim$ 2 Hz.
\cite{dutta2016,dutta2018} concluded that time-lag is a resultant effect of multiple physical mechanisms which are repeated Compton scattering (produces hard-lags), reflection of hard X-rays (soft-lag), focusing due to gravitational bending of photon path (soft-lag) and geometry (Reverberation mapping, produces both soft and hard-lag) of accretion dynamics. These mechanisms are responsible for producing the lags at different Fourier frequencies where time-lag decreases with the increase of the frequency and vice versa. \cite{dutta2016}
reported that the GX 339-4 (face-on) shows hard-lag for all QPO frequencies, whereas for edge-on sources e.g. XTE J1550-564 and GRS 1915+105, time-lag switches sign from positive to negative at QPO frequency $\sim$ 3.2 Hz and $\sim$ 2.3 Hz respectively. Despite all of these developments for lag study at LFQPOs \citep{patra2019}, phase/time lag has not been studied extensively for HFQPOs as it is rare to be observed. GRS 1915+105 shows prominent stable HFQPO at $\sim$ 67 Hz for almost 25 years as observed by both \textit{RXTE} and \textit{AstroSat} missions. Thus, it gives us a unique opportunity to study time-lag features at HFQPOs observed with the \textit{AstroSat}.

GRS 1915+105, a bright microquasar \citep{mirabel1994} of mass $12.4M_{\sun}$ is a rapidly rotating (spin > 0.98) black hole X-ray binary (BH-XRB) source \citep[see][and references therein]{zhang1997,sreehari2020} at a distance of $8.6$ kpc \citep{reid2014} with a jet axis at an angle $\sim$ 65$^{\circ}$ to the line of sight \citep{zdziarski2014}. 
Depending on its variability of lightcurve and color-color diagram (CCD), all the observations of this source can be classified into 14 distinct classes \citep{belloni2000,klein-wolt2002,hannikainen2005}. 
The source variation in each of these classes could be reduced to the alternations of three basic spectral states: A, B, and C. A quiescent state (C) with a lower count rate, an outburst state (B) with higher count rates, and a flare state (A) in which the flux shows rapid changes between the later two states \citep{belloni2000}.
HFQPO was studied extensively using \textit{RXTE} observations with the first detection of HFQPO at 65$-$67 Hz \citep{morgan1997}. 
\cite{belloni2013} detected 63$-$71 Hz HFQPO in seven ($\kappa$, $\gamma$, $\mu$, $\delta$, $\omega$, $\rho$ and $\nu$) variability classes. \cite{cui1999} first attempted to study the lag properties of HFQPO using \textit{RXTE} data and they found that phase-lag increases with energy. \cite{mendez2013} found that 35 Hz QPO shows soft-lag, whereas the 67 Hz QPO exhibits hard-lag and the magnitude of both lag increases with the energy using \textit{RXTE} data.  
The HFQPO of this source was found to be associated with state B, which corresponds to the disc \citep[and references therein]{belloni2013} dominated energy spectrum.
The spectral analysis during this state indicates that the inner radius of the accretion disc can reach close towards the black hole. There was no detection of HFQPO during state C (equivalent to the hard-intermediate state) of the source, instead strong type-C QPOs were observed.

In the \textit{AstroSat} era, a few works have been done on the HFQPOs of GRS 1915+105. \cite{belloni2019} observed HFQPOs in the range of 67.4$-$72.3 Hz considering observations from July to September 2017 observations. They found hard-lag at HFQPO with respect to 5$-$10 keV. \cite{sreehari2020} observed $\delta$ class observations and found that HFQPOs are only present in the 6$-$25 keV energy range and inferred that the HFQPOs in GRS 1915+105 seem to originate due to an oscillation of Comptonized `compact' corona surrounding the central source. Further, \cite{seshadri2022} also studied all observations of this source extensively using \textit{AstroSat} and observed that the HFQPOs are significant in 6$-$25 keV and present in the $\delta$, $\kappa$, $\omega$, $\gamma$ classes having frequency range $68.14-72.32$ Hz. They also concluded that the HFQPOs might be the result of the modulation of the Comptonizing corona in the vicinity of the source. However, the cross-spectrum analysis of those observations was not carried out for calculating the time-lag at the HFQPOs. The study of time-lag properties can be useful for a better understanding of generation of HFQPOs and the accretion flow around the BH-XRBs.

In this paper, for the first time to the best of our knowledge, we are presenting the class-wise time-lag study of the HFQPOs of GRS 1915+105 observed with \textit{AstroSat}. We consider the HFQPO observations as reported by \cite{seshadri2022}. Lightcurve and color-color diagrams were produced to identify the variability classes for those observations. We study the PDS to obtain the HFQPO parameters for further lag study. Lag spectra of the 6$-$25 keV band with respect to the 3$-$6 keV band is produced to find the time-lag feature at HFQPO.
Further, we analyze the energy-dependent time-lag with respect to the 3$-$6 keV band and carried out spectral analysis to correlate the time-lag features with spectral parameters.

We organize this paper as follows. In \S \ref{section:observation}, we present the reduction procedures of \textit{AstroSat} data. In \S \ref{section: analysis}, we discuss the method of timing and spectral analysis and present the results. We discuss the results and attempt to explain them with the existing models present in the literature in \S \ref{section: discussion}. Finally, we make our conclusion in \S \ref{section: conclusions}.

\section{Observation And Data Reduction} \label{section:observation}

We consider all the \textit{AstroSat} observations of HFQPOs of GRS 1915+105 from April 2016 to September 2017 as reported by \cite{seshadri2022}. These observations consist of total 9 Guaranteed Time (GT) observations containing over 70 orbits. Among these, we consider 15 orbits such that they correspond to the observation of each consecutive day. Table \ref{tab:pds_table} summarises the list of considered observation IDs along with orbit number. The maximum exposure time available for all observations is considered for our analysis. We obtain all of the \textit{AstroSat} data from the Indian Space Science Data Center (ISSDC) archive\footnote{\label{issdc}\url{https://webapps.issdc.gov.in/astro_archive/archive/Home.jsp}}.

\textit{AstroSat} \citep{agrawal2006} is India’s first multi-wavelength satellite which was launched in 2015. It consists of four co-aligned instruments which are capable of studying astrophysical objects in the ultra-violet, soft X-ray and also in the hard X-ray regime. These four instruments are Ultra-Violet Imaging Telescope (\textit{UVIT}) \citep{tandon2017}, Soft X-ray Telescope (\textit{SXT}) \citep{singh2017}, Large Area X-ray Proportional Counter (\textit{LAXPC}) \citep{yadav2016,antia2017} and Cadmium Zinc Telluride Imager (\textit{CZTI}) \citep{vadawale2016}. We have used \textit{SXT} and \textit{LAXPC} observations for our study. The timing resolution of 10 $\mu$s in \textit{LAXPC} \citep{yadav2016} and the unique combination of the energy range of both \textit{SXT} and \textit{LAXPC} enable us to study the broadband (0.3$-$80 keV) spectro-temporal feature of the source. 

\textit{SXT} observes the X-ray sources in the energy range 0.3$-$8 keV. We have considered the level-2 data for analysing only the spectra of \textit{SXT} due to the poor time resolution (2.3775 s) compared to \textit{LAXPC}. The source image, lightcurve, and spectra are generated by using \texttt{XSELECT V2.4m} in \texttt{HEASOFT V6.29c}. The source has been observed by \textit{SXT} both in Photon Count (PC) and Fast Window (FW) mode.
We do not carry out pile-up correction for the observations in PC mode as the source counts are less than 40 cts/s in the central 1 arcmin circular region of the image \citep{seshadri2022}. For all PC mode observations, a circular region of 12 arcmins is selected for extracting the spectra.
In the case of the FW mode observations, we choose the 5 arcmins circular region as the source region as it was offset in the CCD frame \citep{sreehari2020}. We use the background file, redistribution matrix files (rmfs), and ancillary response files (arfs) provided by the \textit{SXT} instrument team at TIFR\footnote{\url{https://www.tifr.res.in/~astrosat_sxt/dataanalysis.html}}. Further, we use \texttt{sxtARFModule} to apply vignetting correction on the provided arf file.

\textit{LAXPC} is a proportional counter which can detect X-rays in the energy range 3$-$80 keV. It consists of three identical unit, \textit{LAXPC10}, \textit{LAXPC20} and \textit{LAXPC30} having combined effective area of 6000 cm$^{2}$. We use \textit{LAXPC} level-1 data available in \textit{AstroSat} public archive$^{\ref{issdc}}$. The software \texttt{LAXPCsoftware}, available in \textit{AstroSat} science support cell\footnote{\url{http://astrosat-ssc.iucaa.in}} is used to process the level-1 data to level-2 data. This level-2 data is used for further timing analysis. We extract data from all layers of \textit{LAXPC10} and \textit{LAXPC20} for the temporal analysis. However, we use \texttt{LaxpcSoftv3.4.3} \citep{antia2017} for the spectral analysis of \textit{LAXPC} observations. We generate the Good time Interval (GTI) and consider the background model and spectral response as mentioned in \cite{athulya2022}. In order to minimize the residue in the spectra beyond 30 keV, we extract the top layer, single event data from \textit{LAXPC20} because of its steady gain throughout the considered observational period. We didn't consider \textit{LAXPC30} because its gain was observed to increase with time due to a leakage \citep{antia2021}.

\section{Analysis and Results} \label{section: analysis}

\subsection{Variability Classes and Color-Color Diagram (CCD)} \label{subsec:lc_ccd}

 GRS 1915+105 exhibits different variability features in its light curve and Color-Color Diagrams (CCDs). The source displays fourteen variability classes \citep{belloni2000,klein-wolt2002,hannikainen2005} as observed by the \textit{RXTE}. 
 CCDs are used to understand the hardness variation during different spectral states and to identify the nature of the variability class exhibited by the source.
In order to produce CCDs, we generate the 1 s binned lightcurves in the energy range 3$-$6 keV, 6$-$15 keV and 15$-$60 keV by combining \textit{LAXPC10} and \textit{LAXPC20}. We perform background correction to these lightcurves and generate CCD to study the variation between soft color HR1 and hard color HR2. The soft color and hard color are defined as HR1=B/A and HR2=C/A respectively, where A, B, and C are the count rates in 3$-$6 keV, 6$-$15 keV and 15$-$60 keV respectively. Finally, we add A, B and C bands lightcurve to produce the lightcurve in 3$-$60 keV band and perform dead-time corrections \citep[and references therein]{sreehari2020}.
However, the power density spectra are generated from the lightcurve without any background correction.

\begin{figure}   
 \includegraphics[width=\columnwidth]{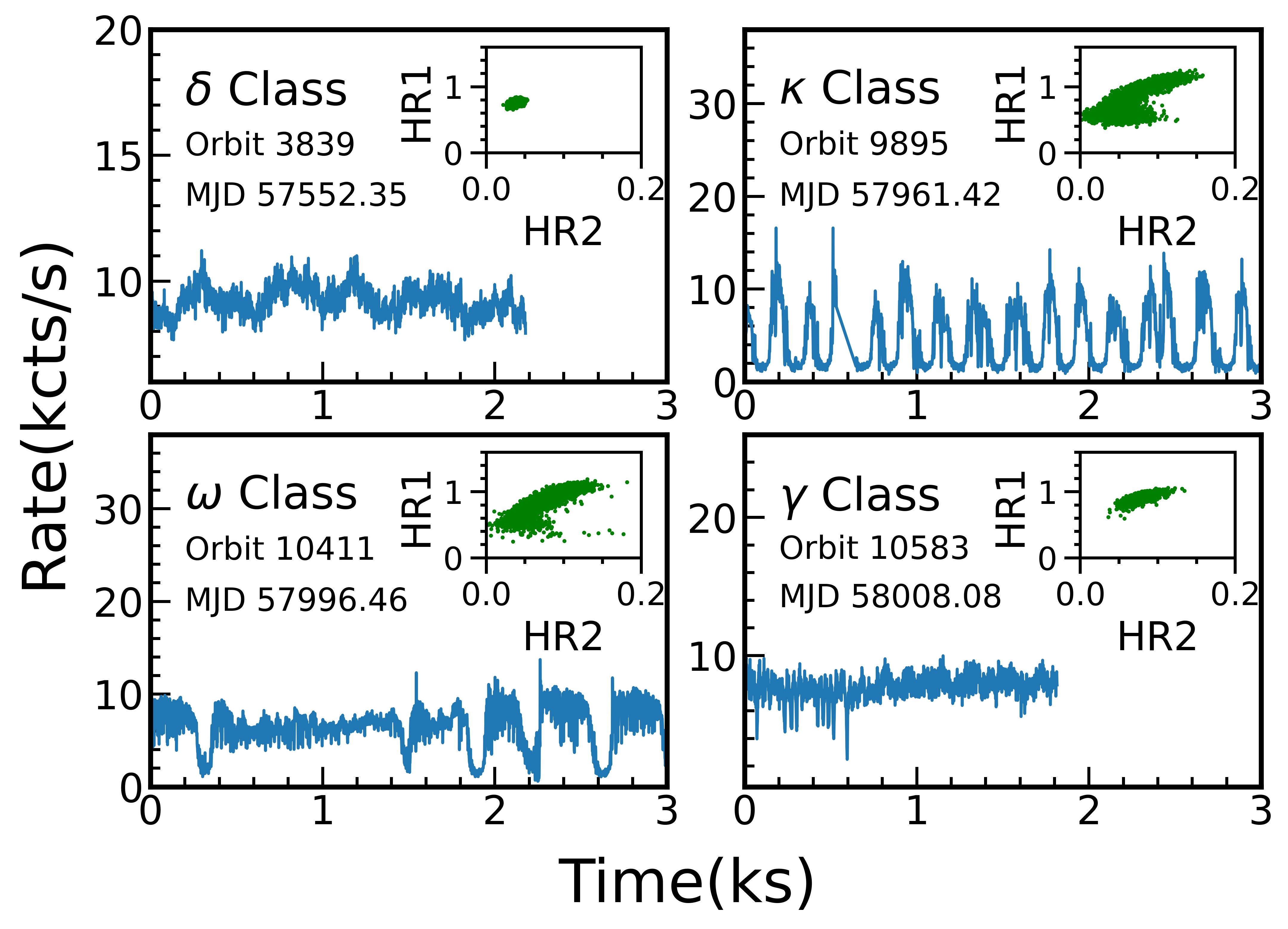}
 \caption{Lightcurve and CCD of the source GRS 1915+105 during HFQPO observations using \textit{AstroSat}. The background subtracted and dead-time corrected 1 s binned \textit{LAXPC} lightcurve corresponding to four variability classes namely $\delta$, $\kappa$, $\omega$, and $\gamma$ are plotted in the 3$-$60 keV energy range with CCD (top right inset). See text for detail.}
 \label{fig:lc_ccd_plot}
\end{figure}

Following the classification scheme in \cite{athulya2022} and \cite{seshadri2022}, we identify that the considered \textit{AstroSat} observations of HFQPOs belong to the four variability classes $\omega$, $\delta$, $\kappa$ and $\gamma$. The background subtracted and dead-time corrected lightcurves of four classes are shown in Fig. \ref{fig:lc_ccd_plot} with CCD in the inset at the top right corner of each panel.

\subsection{HFQPOs in PDS and RMS Power}\label{subsec:pds}

\begin{figure}
 \includegraphics[width=\columnwidth]{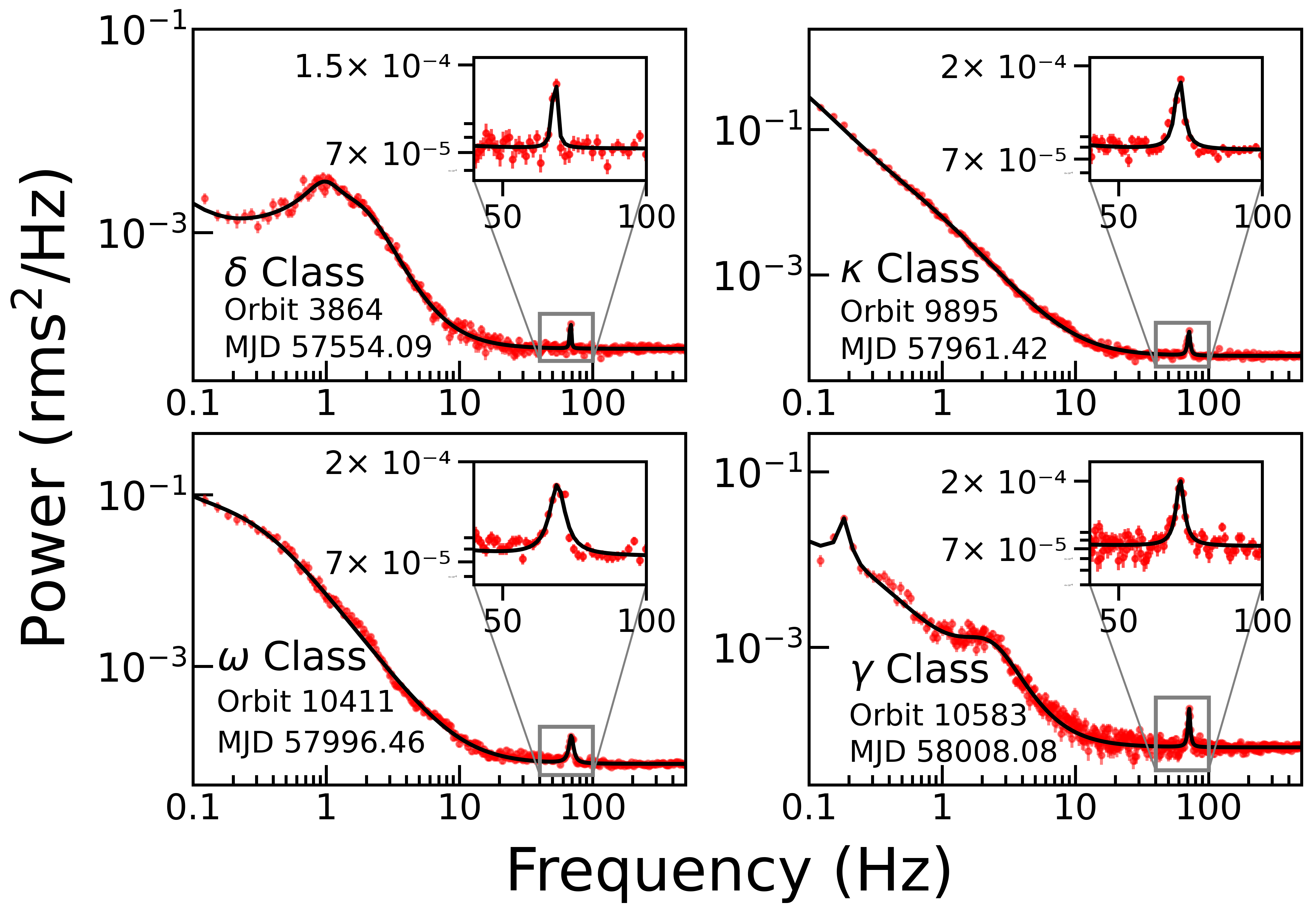}
 \caption{Power density spectra generated from 1 ms binned lightcurve for four variability classes ($\delta$, $\kappa$, $\omega$, $\gamma$) in 3$-$60 keV energy range. In each panel variability class along with orbit number is mentioned, and the HFQPO feature (zoomed view) is also shown in the upper right inset. See text for detail.}
 \label{fig:pds_plot}
\end{figure}

We limit our analysis to those observations where only HFQPOs are observed. We produce the PDS of all considered observations from the four variability classes (see \S\ref{subsec:lc_ccd} and Table \ref{tab:pds_table}) in order to calculate the HFQPO parameters for further cross-spectral study. We generate 1 ms binned lightcurve in 3$-$60 keV band to produce a PDS with Nyquist frequency of 500 Hz. We consider intervals of 32.768 s and compute the Fast Fourier Transform (FFT) for each interval, and we combined them using a geometric binning factor of 1.02 to get an averaged PDS. The PDS are standardized
and converted into the Leahy power spectrum \citep{leahy1983}.
For each observation, we subtracted the Poisson noise contribution \citep{zhang1995,yadav2016} and also corrected the dead-time effects on rms amplitude \citep{bachetti2015,sreehari2019} to produce the final PDS in \textit{rms$^{2}$/Hz} plane.
The power spectra were then fitted with a combination of \texttt{constant} and multiple \texttt{Lorentzian} \citep[see][]{nowak2000}.
The PDS are shown for all four classes in Fig. \ref{fig:pds_plot} with the HFQPOs in the inset at the upper right corner. The variability class along with the orbit are mentioned in the lower right corner of each panel. 

In order to obtain the HFQPO parameters only, we further fit the PDS in the frequency range of 20$-$200 Hz. We fit all the PDS by using a \texttt{constant} and a \texttt{Lorentzian} feature with centroid frequency at $\sim$ 68 Hz. This model yields the reduced $\chi^{2}$ ($\chi^{2}_{red}=\chi^{2}/dof$) in the range 1.0$-$1.3 for all PDS which ensures the satisfactory fit. The best-fitted values of centroid frequency, FWHM and percentage rms amplitude are produced in Table \ref{tab:pds_table}. The percentage rms is calculated by finding the square root of the definite integral of the fitted \texttt{Lorentzian} and multiplying it by 100 \citep[see][]{athulya2022}. We calculate the error in rms by using the error propagation rule. We use centroid frequency and FWHM for the calculation of time-lag at the HFQPOs.


\begin{table*}

    \centering
    \caption{\label{tab:pds_table}Details of the best fitted PDS parameters from \textit{LAXPC} observations of GRS 1915+105 in 3$-$60 keV energy range for four variability classes are presented. Results are obtained after combining the \textit{LAXPC10} and \textit{LAXPC20}. All PDS are fitted in the frequency range 20$-$200 Hz in rms$^{2}$/Hz plane. Here, in this table, model fitted parameters of HFQPO (centroid frequency and FWHM) are given along with percentage rms amplitude. The errors of fitted parameters are computed with 68\% confidence range. Time-lags are calculated for 6$-$25 keV with respect to 3$-$6 keV and tabulated in units of milliseconds. See text for details.}
    \begin{tabular}{ccccccccccc}
	\hline
    \hline
Obs ID & MJD & Orbit & HFQPO$_{freq}$ & FWHM & 
HFQPO$_{rms}$\% & Time-lag & Class \\
 &   &       & (Hz)           & (Hz) &  & (ms) &    \\
    \hline
    \hline
G05\_214T01\_9000000428 & 57503.63 & 03117 & 70.77$^{+0.11}_{-0.11}$ & 1.72* & 
1.24$\pm$0.07\%  & -0.51$\pm$ 0.15 & $\omega$ \\

   &  57504.02 & 03124 & 70.78$^{+0.24}_{-0.30}$ & 3.64$^{+1.36}_{-1.35}$ & 
   1.34$\pm$0.28\%  & -0.40$\pm$0.07 & $\omega$     \\

G05\_189T01\_9000000492 & 57551.04 & 03819 & 68.02$^{+0.06}_{-0.05}$ & 3.00* & 
1.57$\pm$0.05\% & -1.06$\pm$0.17 & $\delta$  \\

&  57552.35 & 03839 & 68.42$^{+0.45}_{-0.23}$ & 1.46* & 
0.74$\pm$0.10\%  & -0.49$\pm$0.16 & $\delta$  \\

&  57553.88 & 03860 & 68.13$^{+0.20}_{-0.24}$ & 2.93$^{+0.70}_{-0.81}$ &
1.40$\pm$0.19\%   & -0.74$\pm$0.12    & $\delta$   \\

&  57554.09 & 03864 & 68.14$^{+0.10}_{-0.09}$ & 3.00* & 
1.49$\pm$0.07\%  & -1.00$\pm$0.16 & $\delta$  \\

G07\_028T01\_9000001370 & 57943.69 & 09633 & 70.67$^{+0.45}_{-0.46}$ & 4.25$^{+1.48}_{-1.42}$ &
1.77$\pm$0.35\%     & -0.54$\pm$0.22 &  $\kappa$   \\

G07\_046T01\_9000001374 & 57945.97 & 09667 & 71.14$^{+0.11}_{-0.11}$ & 2.83$^{+0.48}_{-0.41}$ &
2.01$\pm$0.17\%    & -0.80$\pm$0.21   & $\kappa$       \\

&  57946.34 & 09670 & 70.30$^{+0.22}_{-0.23}$ & 5.10$^{+0.54}_{-0.48}$ &
2.50$\pm$0.15\%    & -0.91$\pm$0.18   & $\kappa$    \\

G07\_028T01\_9000001406 & 57961.42 & 09895 & 71.16$^{+0.08}_{-0.08}$ & 2.92$^{+0.31}_{-0.29}$ &
2.09$\pm$0.12\%    & -0.79$\pm$0.13   & $\kappa$    \\

G07\_046T01\_9000001408 & 57962.07 & 09902 & 70.03$^{+0.28}_{-0.28}$ & 5.66$^{+0.86}_{-0.74}$ &
2.58$\pm$0.21\%   & -0.81$\pm$0.21   & $\kappa$       \\

G07\_028T01\_9000001500 & 57995.30 & 10394 & 67.85$^{+0.17}_{-0.19}$ & 4.17$^{+0.47}_{-0.51}$ &
2.20$\pm$0.16\%   & -1.50$\pm$0.15   & $\omega$      \\

G07\_046T01\_9000001506 & 57996.46 & 10411 & 69.10$^{+0.13}_{-0.14}$ & 4.92$^{+0.31}_{-0.31}$ & 
2.46$\pm$0.09\%   & -1.07$\pm$0.09   & $\omega$   \\

G07\_046T01\_9000001534 & 58007.59 & 10575 & 71.62$^{+0.14}_{-0.14}$ & 4.04$^{+0.40}_{-0.35}$ & 
2.07$\pm$0.11\%   & -0.88$\pm$0.11   & $\omega$  \\

&  58008.08 & 10583 & 71.41$^{+0.08}_{-0.08}$ & 2.70$^{+0.26}_{-0.23}$ & 
2.32$\pm$0.12\%   & -1.68$\pm$0.20   & $\gamma$  \\
	\hline
	\end{tabular}
    \begin{list}{}{}
		\item[*] Fixed parameter.
	\end{list}
    
\end{table*} 


\subsection{Time Lag properties of HFQPO} \label{subsec:time-lag}

\subsubsection{Lag-Spectra} \label{subsubsec:lag-spectra}

X-ray lags are measured using the cross 
Fourier spectrum of X-ray light curves with different energy
bands. The soft energy band is chosen on the basis of the detector's lower energy limit and also on the nature of the source.
The harder photons, energized via inverse Compton process \citep{sunyaev_titarchuk1980}, are used to cross-correlate with the soft photons to generate frequency dependent time-lag spectra.

\begin{figure*}
 \includegraphics[width=0.7\textwidth]{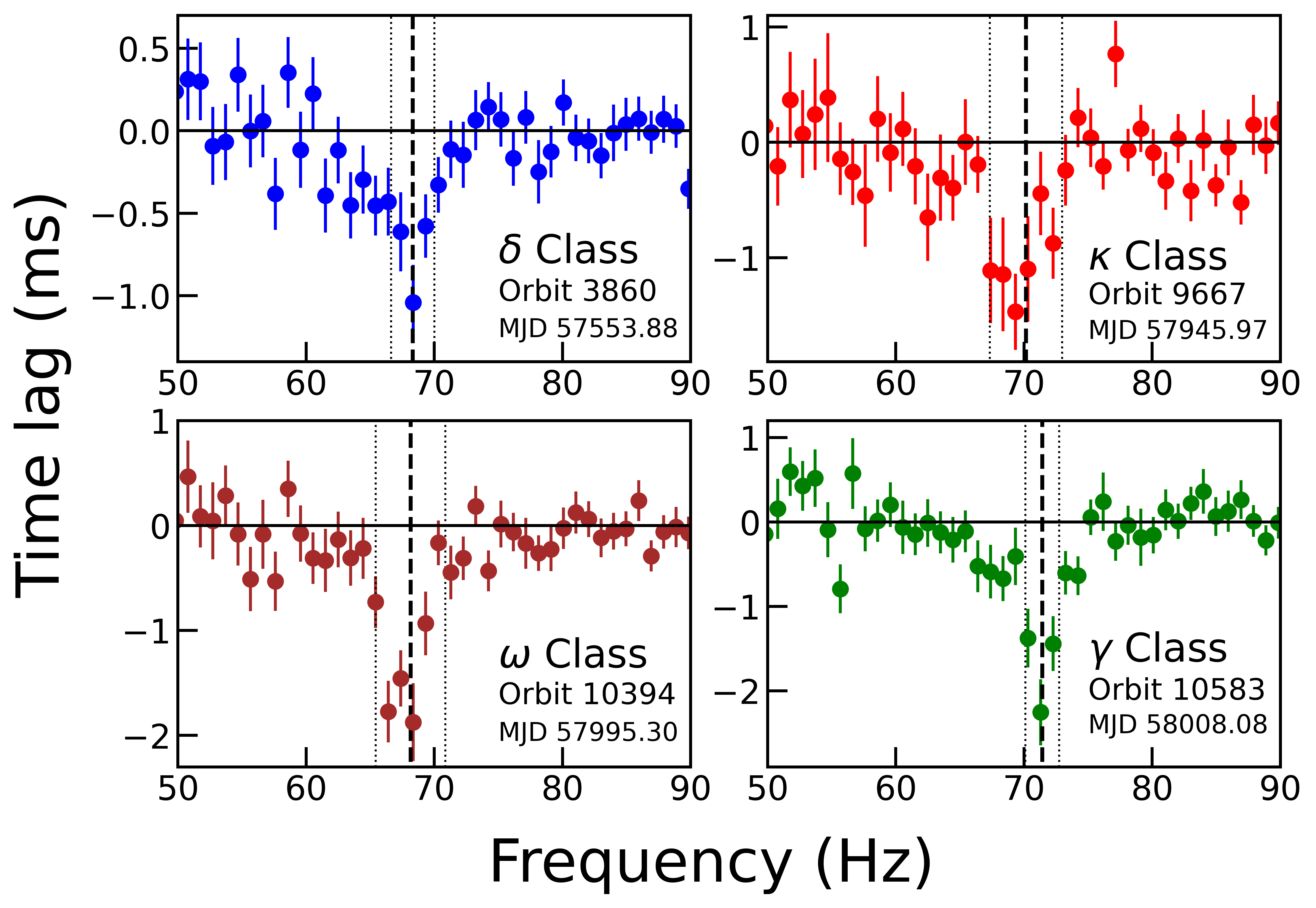}
 \caption{Lag spectra (time-lag as a function of Fourier frequency) is plotted for photons in the energy band 6$-$25 keV with respect to those in 3$-$6 keV band for four variability classes. All variability classes show soft-lag at HFQPO. The centroid frequency of HFQPO is shown as the vertical black dashed line and the vertical black dotted lines represent the range of FWHM of the HFQPO.}
 \label{fig:lag-spectra_plot}
\end{figure*}

We generate time-lag spectra for each observation. Time-lag between the photons of two different energy bands can be calculated from the argument of cross spectra. 
Let, the photon count in two different simultaneous energy bands is observed to be $x_{1}(k)$ and $x_{2}(k)$ at time $t_{k}$ and their Fourier transforms are $X_{1}(j)$ and $X_{2}(j)$ respectively at frequency $\nu_{j}$. The Fourier transform is defined as,
\setlength{\abovedisplayskip}{5pt}
\setlength{\belowdisplayskip}{5pt}
\begin{equation}
    X(j) = \sum_{k=0}^{m-1} x(k)e^{2\pi i \nu_{j} t_{k}}.
\end{equation}
Here, $x(k)$ is the count in the $k$th bin of the lightcurve which consists of $m$ evenly spaced time-bin of length $\Delta t$ in seconds. $X(j)$ is the complex Fourier transform of $x(k)$ corresponding to frequency $\nu_{j}=j/(m\Delta t)$ where, $j \in [-m/2, m/2]$ \citep[see][for details]{van_der_Klis1989_Fourier_technic}.
Cross spectra are defined as, $C(j)=X^{*}_{1}(j)X_{2}(j)$
corresponding to Fourier frequency $\nu_{j}$. $X^{*}_{1}(j)$ refers to the complex conjugate of $X_{1}(j)$.
The position angle of $C(j)$ in complex plane is defined as phase-lag, $\phi(j)=arg[C(j)]$ and the corresponding time-lag is defined as, $\delta t(j)=\phi(j)/2\pi\nu_{j}$ \citep[see][for details]{vaughan_nowak1997}. The variation of this time-lag $\delta t(j)$ as a function of Fourier frequency $\nu_{j}$ is known as time-lag spectra. Following the methodology mentioned above, we divide each observation into two energy bands, 3$-$6 keV and 6$-$25 keV and produce the time-lag of 6$-$25 keV photons as a function of Fourier frequency with respect to 3$-$6 keV photons. We consider Nyquist frequency of lag spectra to be 500 Hz in order to perform time-lag analysis near HFQPO$_{freq}$ (see \S \ref{subsec:pds}) which is well below the region where binning effects are important \citep[see][for more details]{crary1998}. The Fourier frequency is binned at an interval $\sim$ 1 Hz as it offers much less error on time-lag. We did not perform any dead-time correction on lag-spectra since this effect was found to be negligible \citep{bachetti2015}. The lag-spectra of all four variability classes are shown in Fig. \ref{fig:lag-spectra_plot}. The variability class and the orbit are mentioned in the bottom right corner of each panel. The vertical black dashed line shows the HFQPO$_{freq}$ and the vertical black dotted lines represent the FWHM range of the HFQPOs. The negative lag means soft-lag where soft photons (i.e., 3$-$6 keV) are lagging behind the hard photons (i.e., 6$-$25 keV). All the variability classes exhibit a similar coherent soft-lag feature of different magnitudes associated with the HFQPOs.

We calculate the time-lag from this lag-spectra following \cite{reig2000} by averaging the time-lag over the frequency range ($\nu_{QPO} - \Delta\nu_{QPO}/2$) to ($\nu_{QPO} + \Delta\nu_{QPO}/2$), where $\nu_{QPO}$ and $\Delta\nu_{QPO}$ is the centroid frequency and FWHM of \texttt{Lorentzian} respectively, obtained from the fitted PDS in \S\ref{subsec:pds}. We calculate the time-lag from each observation and presented in Table \ref{tab:pds_table}. Among all classes, the $\kappa$ variability class shows the soft-lag in the range of 0.54$-$0.91 ms. The only observation of the $\gamma$ class shows the maximum soft-lag of 1.68 ms. The $\delta$ and $\omega$ classes show a wide range of time-lag. The $\delta$ class exhibits soft-lag in the range 0.49$-$1.06 ms, whereas the soft-lag of the $\omega$ class lies in the range of 0.40$-$1.50 ms.

\begin{figure*}
 \includegraphics[width=0.7\textwidth]{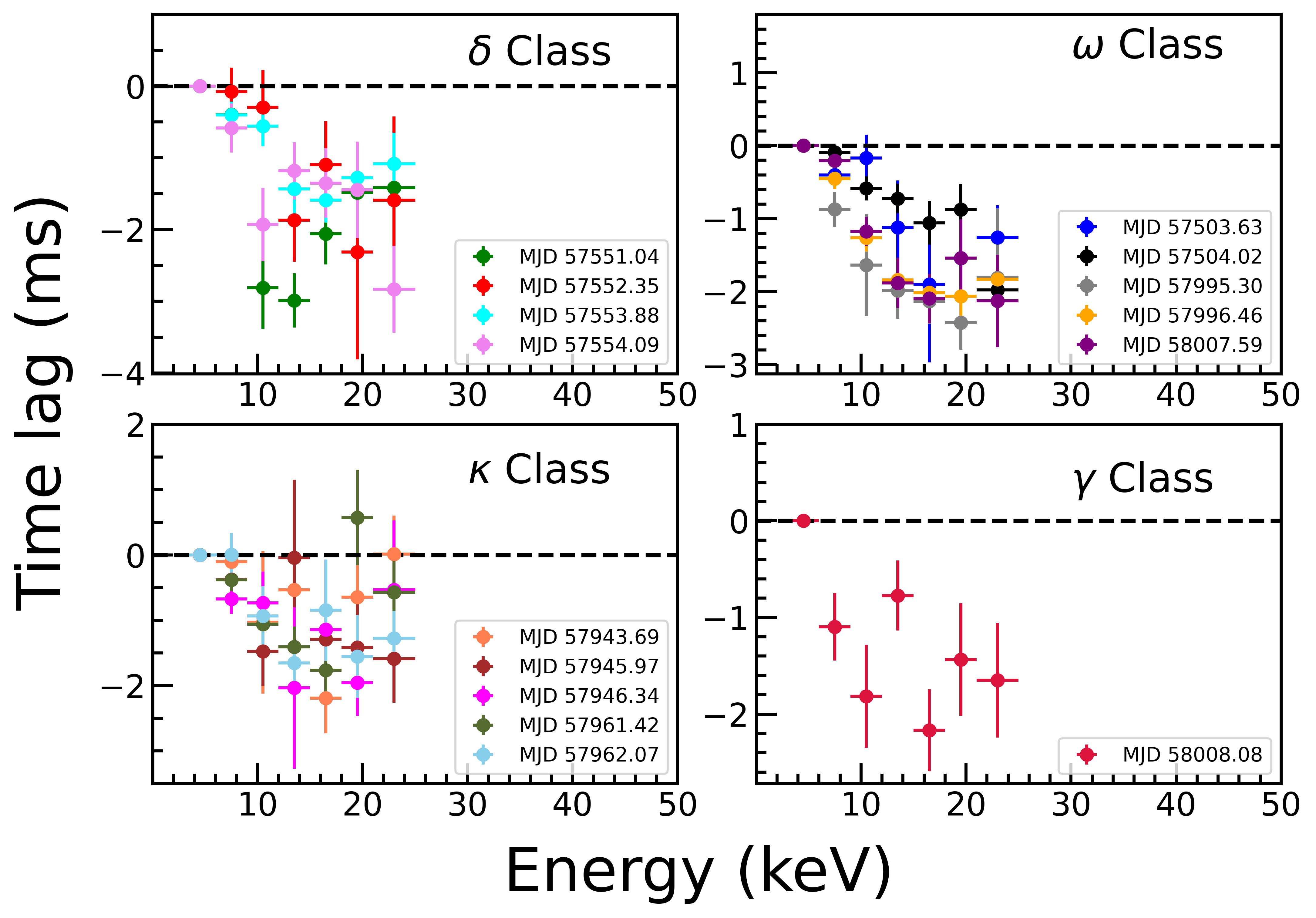}
 \caption{Energy-dependent time lag w.r.t 3$-$6 keV band at the HFQPO frequency is plotted for the four variability class observations. Variability classes and the orbits of all observations are mentioned in the upper right and lower right corners of each panel respectively. See text for details.}
 \label{fig:lag-energy}
\end{figure*}

\subsubsection{Energy Dependent Time-lag} \label{subsubsec:lag_energy}

Time-lag is a manifestation of different non-linear mechanisms. These mechanisms produce different energy photons at different regions of the accretion flow. Thus, the observed energy-dependent time-lag feature can explain the responsible mechanism for producing different energy photons. 
Time-lag of any particular energy band depends on the respective reference energy band.
In order to study the time-lag variations with energy, we divide the entire energy band 3$-$25 keV into seven different energy bands. The chosen energy bands are: 3$-$6 keV, 6$-$9 keV, 9$-$12 keV, 12$-$15 keV, 15$-$18 keV, 18$-$21 keV and 21$-$25 keV. We generate lag-spectra for each energy band with respect to the softest energy band 3$-$6 keV. Lag-spectra is produced in a similar way as mentioned in the previous subsection \S \ref{subsubsec:lag-spectra}. From each lag-spectra, we calculate the average time-lag over the FWHM of the HFQPO (see Table \ref{tab:pds_table}) and we plotted them with respect to their corresponding energy.

The time-lag as a function of energy for all four variability classes and for all orbits are plotted in Fig. \ref{fig:lag-energy}. The variability class and the MJD are mentioned in the upper left and lower left corners respectively of each panel. It is clear from the figure that the time-lag at the HFQPO depends strongly on photon energy. We find for all of the observations that the magnitude of the soft-lag increases with energy up to $\sim$ 18 keV. However, different observations in each class show different time-lag variations with energy. In the $\delta$ variability class, maximum soft-lag of $\sim$ 3 ms is observed on MJD 57551.04 (orbit 3819), whereas among the $\omega$ class observations, maximum soft-lag of $\sim$ 2.5 ms on MJD 57995.30 (orbit 10394) is observed. Moreover, in the $\kappa$ and $\gamma$ variability classes, maximum soft-lag of $\sim$ 2.1 ms is observed on MJD 57943.69 (orbit 9633) and MJD 58008.08 (orbit 10583). It may be noted that we do not encounter the problem of dead-time-driven cross-talk. In this problem, a $+\pi$ or $-\pi$ phase-lag is produced in the frequency domain where Poisson noise is dominated \citep[see][]{van_der_klis1987,mendez2013}.

\subsection{Spectral Analysis and Results} \label{subsec:spectral_analysis}

In order to correlate the time-lag with spectral parameters, we perform wide band (0.7$-$50 keV) spectral analysis from simultaneous data of \textit{SXT} and \textit{LAXPC} for each variability class \citep[see][for details]{seshadri2022}. \textit{SXT} spectra are extracted in the energy range 0.7$-$7.0 keV and \textit{LAXPC} spectra are extracted in the energy range 3$-$50 keV. \textit{SXT} spectra are grouped with 30 counts in each bin before modelling, whereas grouping is not applied for \textit{LAXPC20} spectra. We ignore spectral analysis for some of the observations due to the poor quality of \textit{SXT} data. 

Both \textit{SXT} and \textit{LAXPC20} spectra are modelled using the \texttt{XSPEC V12.12.0} which is available in \texttt{HEASOFT V6.29.c}. A systematic error of 2\% is considered for both spectra \citep{antia2017,leahy_chen2019,sreehari2019}. We considered \texttt{gain fit} command in \textit{SXT} spectra to alter the response to fit the edges at 1.8 keV and 2.2 keV produced due to the Si and Au \citep{singh2017}. During the application of \texttt{gain fit}, we fix the slope at 1 and allow the offset to vary. 

\begin{figure}
 \includegraphics[width=\columnwidth]{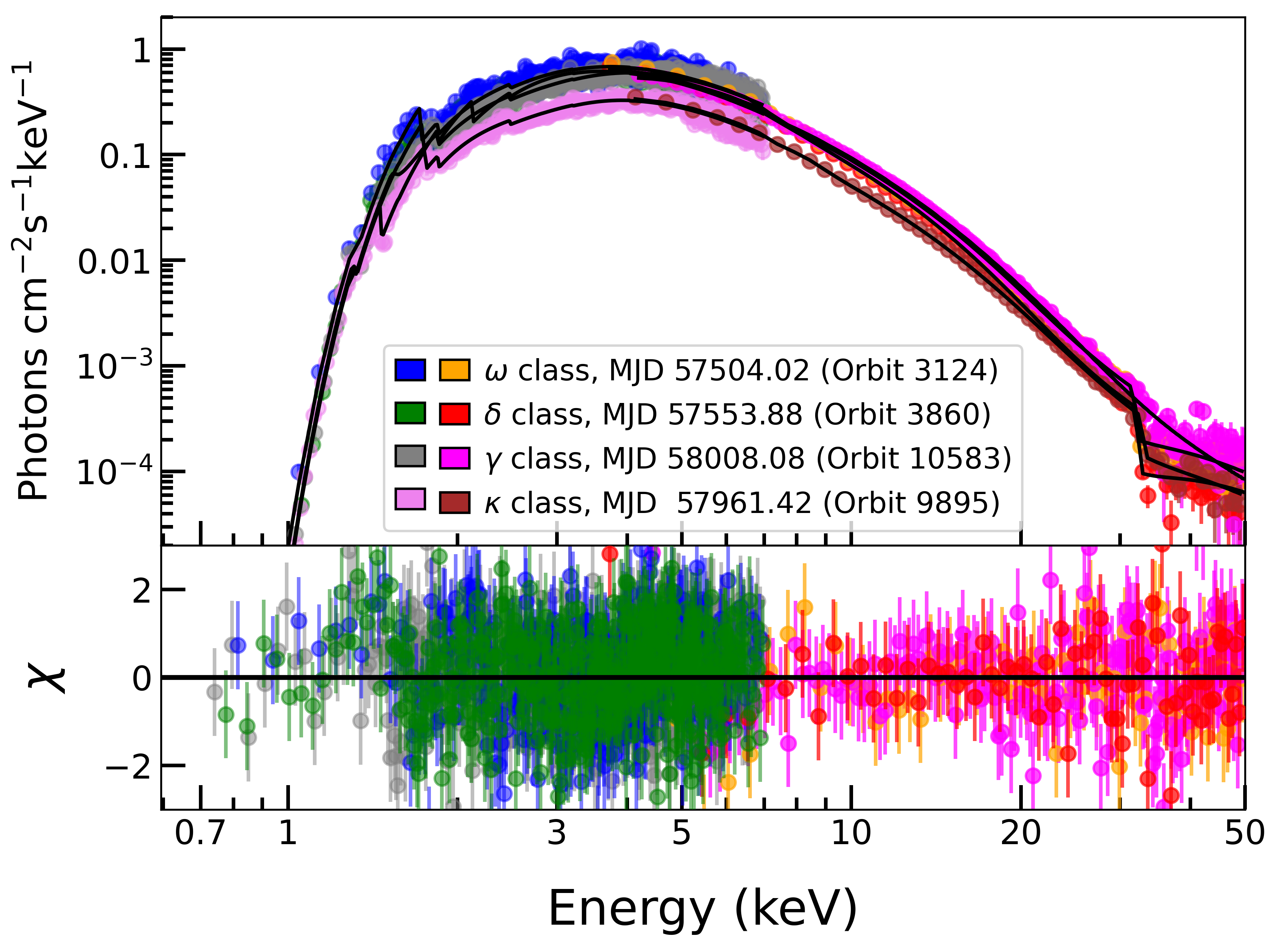}
 \caption{Wide band (0.7$-$50.0 keV) spectra of the source GRS 1915+105 are plotted for four variability classes. The spectra are modelled with \texttt{constant$\times$TBabs$\times$(smedge$\times$nthComp + edge$\times$powerlaw)}. The variability classes and the orbits are mentioned.}
 \label{fig:spectrum}
\end{figure}

We fitted all the spectra with the model combination of \texttt{constant$\times$TBabs$\times$(smedge$\times$nthComp + edge$\times$powerlaw)} as discussed in \cite{seshadri2022}.
During the spectral fitting of all observations, we have kept the seed photon temperature ($kT_{bb}$) fixed at 0.1 keV. All the best-fitted model parameters are shown in Table \ref{tab:spectral_parameter}. The instrumental Xenon \texttt{edge} model is used with power law to take care of the instrumental absorption feature at $\sim$ 30 keV in the spectra \citep{sreehari2019}. In order to obtain the best fit, the model component \texttt{smedge} is also required due to a broad absorption feature at $\sim$ 9 keV. The best-fitted wide band spectra of four variability classes are shown in Fig. \ref{fig:spectrum}.
The optical depth ($\tau$) of the Comptonizing medium is obtained using the relation given by \cite{zdziarski1996}.
The calculated optical depth is found to be in the range 6.90$-$12.55, and the HFQPOs are observed during the softer spectral state, which in turn implies the presence of an optically thick Comptonizing region close to the black hole.

\begin{table*}
	\centering
	 \caption{\label{tab:spectral_parameter}Model parameters of the observations fitted with model \text{\textbf{constant$\times$TBabs$\times$(smedge$\times$nthComp + edge$\times$powerlaw)}}. All errors are calculated with 90\% confidence range. }
	 \resizebox{19cm}{!}{
	 \begin{tabular}{cccccccccccccc}
	 \hline
	 \hline

MJD (orbit)  &  $kT_{e}$  &  $\Gamma_{nth}$  & $norm_{nth}$ & $\Gamma_{PL}$ & $norm_{PL}$ & $\chi^{2}_{red}$  & Optical depth &
Class   \\
       & (keV) &    &     &     &   &   & ($\tau$)    \\ 
      
	 \hline
	 \hline 

57504.02 (03124) & 2.59$_{-0.09}^{+0.10}$ & 2.36$_{-0.09}^{+0.09}$ & 38.85$_{-6.63}^{+7.85}$ & 3.26$_{-0.02}^{+0.02}$ & 28.09$_{-0.96}^{+0.87}$ & 1.13 
& 8.60$\pm$0.31  
& $\omega$                   \\

57551.04 (03819) & 2.76$_{-0.14}^{+0.16}$ & 2.28$_{-0.11}^{+0.10}$ & 
30.52$_{-7.15}^{+8.43}$ & 3.16$_{-0.06}^{+0.06}$ & 30.03$_{-4.53}^{+4.00}$ & 1.12 
& 8.68$\pm$0.58
&  $\delta$      \\

57552.35 (03839) & 2.67$_{-0.13}^{+0.16}$ & 2.77$_{-0.12}^{+0.09}$ & 126.39$_{-26.39}^{+28.92}$ & 3.42$_{-0.05}^{+0.04}$ & 46.87$_{-5.39}^{+2.55}$ & 1.07 & 6.90$\pm$0.48  &  $\delta$      \\

57553.88 (03860) & 2.83$_{-0.08}^{+0.06}$ & 2.45$_{-0.06}^{+0.03}$ & 48.88$_{-4.19}^{+4.19}$ & 3.19$_{-0.01}^{+0.01}$ & 32.59$_{-1.51}^{+2.33}$ & 1.11 
& 7.75$\pm$0.28 
&    $\delta$     \\

57554.09 (03864) & 2.85$_{-0.17}^{+0.14}$ & 2.14$_{-0.04}^{+0.08}$ & 19.12$_{-4.01}^{+4.79}$ & 3.04$_{-0.24}^{+0.10}$ & 15.44* & 1.05 & 9.28$\pm$0.62  &  $\delta$      \\

57943.69 (09633) & 2.79$_{-0.11}^{+0.15}$ & 2.55$_{-0.07}^{+0.09}$ & 20.02$_{-3.27}^{+4.22}$ & 3.28$_{-0.03}^{+0.05}$ & 12.81$_{-0.43}^{+0.58}$ & 1.18 
&  7.45$\pm$0.31 
&   $\kappa$     \\

57946.34 (09670) & 2.62$_{-0.07}^{+0.14}$ & 2.33$_{-0.08}^{+0.12}$ &
15.65$_{-2.69}^{+2.31}$ & 3.12$_{-0.01}^{+0.03}$ & 11.56$_{-0.28}^{+0.42}$ & 1.03 
&  8.69$\pm$0.50 
&   $\kappa$     \\

57961.42 (09895) & 2.83$_{-0.07}^{+0.11}$ & 2.27$_{-0.05}^{+0.06}$ &
18.25$_{-2.46}^{+3.40}$ & 3.07$_{-0.02}^{+0.03}$ & 13.15$_{-0.34}^{+0.44}$ & 1.09 
& 8.61$\pm$0.33 
&    $\kappa$    \\

57962.07 (09902) & 2.70* &  2.27$_{-0.04}^{+0.04}$ & 18.37$_{-1.84}^{+2.12}$ & 2.97$_{-0.02}^{+0.02}$ & 12.29$_{-0.49}^{+0.50}$ & 1.02 
& 8.84$\pm$0.21
& $\kappa$ \\

57995.30 (10394) & 2.60* & 2.14$_{-0.05}^{+0.04}$ & 20.33$_{-2.38}^{+2.24}$ & 2.93$_{-0.01}^{+0.01}$ & 17.48$_{-0.47}^{+0.50}$ & 1.02 
& 9.78$\pm$0.31 
& $\omega$  \\

57996.46 (10411) & 2.89$_{-0.11}^{+0.11}$ & 2.27$_{-0.08}^{+0.06}$ &  24.97$_{-3.94}^{+3.90}$ & 3.01$_{-0.03}^{+0.03}$ & 17.09$_{-0.56}^{+0.53}$ & 1.12 
& 8.50$\pm$0.40 
&  $\omega$        \\

58007.59 (10575) & 2.71$_{-0.10}^{+0.11}$ & 2.19$_{-0.07}^{+0.09}$ & 24.47$_{-3.84}^{+5.15}$ & 3.05$_{-0.03}^{+0.03}$ & 19.00$_{-0.62}^{+0.56}$ & 1.05 
& 9.25$\pm$0.62  
& $\omega$   \\

58008.08 (10583) & 2.40* & 1.85$_{-0.04}^{+0.05}$ & 13.45$_{-1.92}^{+2.08}$ & 2.90$_{-0.01}^{+0.01}$ & 19.30$_{-0.71}^{+0.70}$ & 0.95 
& 12.55$\pm$0.48   
&   $\gamma$      \\

	 \hline
	 \end{tabular}
}	 
	 \begin{list}{}{}
		\item[*]Parameter is fixed.
	\end{list}
\end{table*}


\section{Discussion} \label{section: discussion}

\begin{figure}
 \includegraphics[width=\columnwidth]{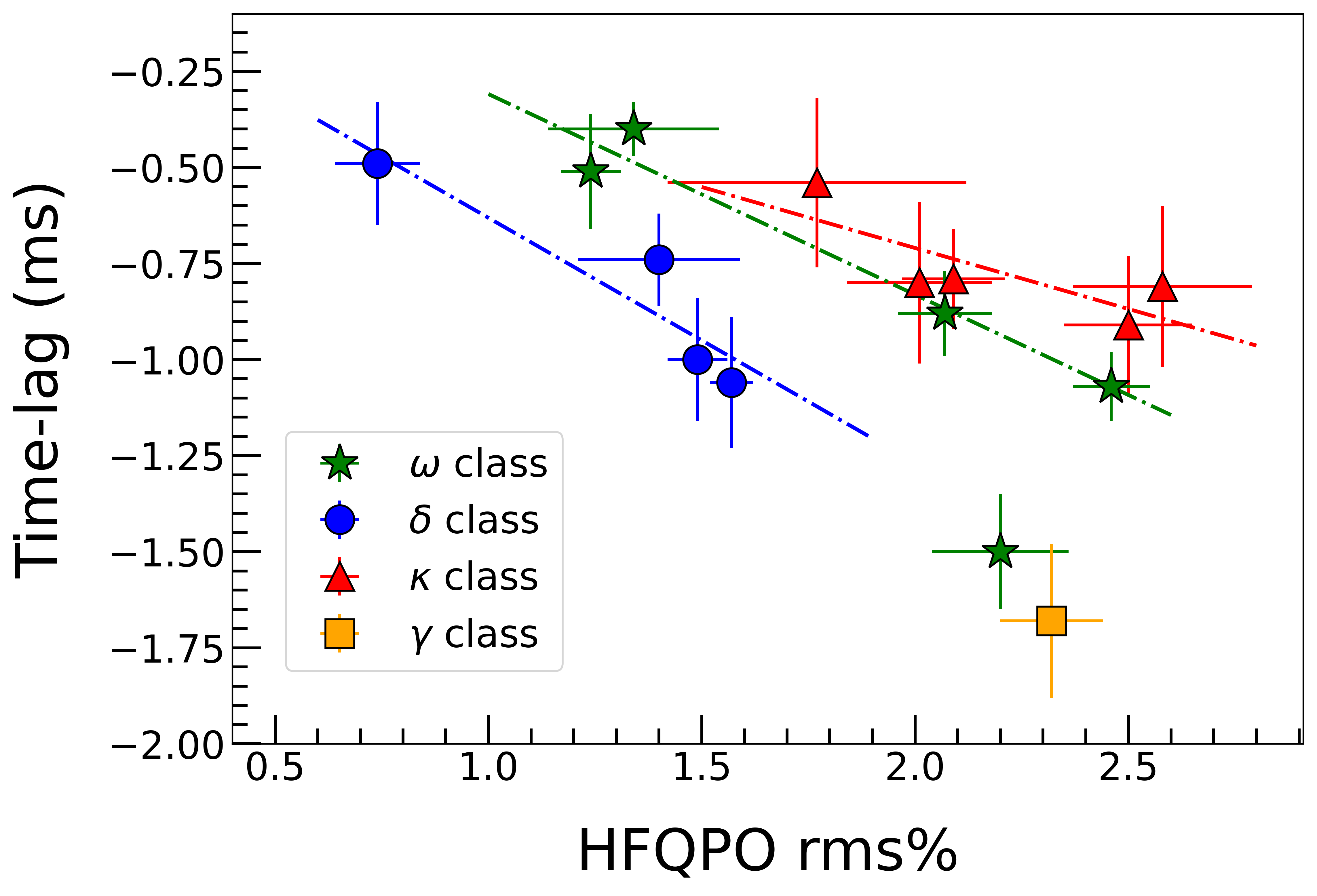}
 \caption{Time-lag of 6$-$25 keV band with respect to 3$-$6 keV band is plotted as a function of rms amplitude of HFQPOs in full energy band. The variability classes are also mentioned in the inset at the lower left corner of the panel. See the text for details.}
 \label{fig:lag_rms_corr}
\end{figure}

\begin{figure}
 \includegraphics[width=\columnwidth]{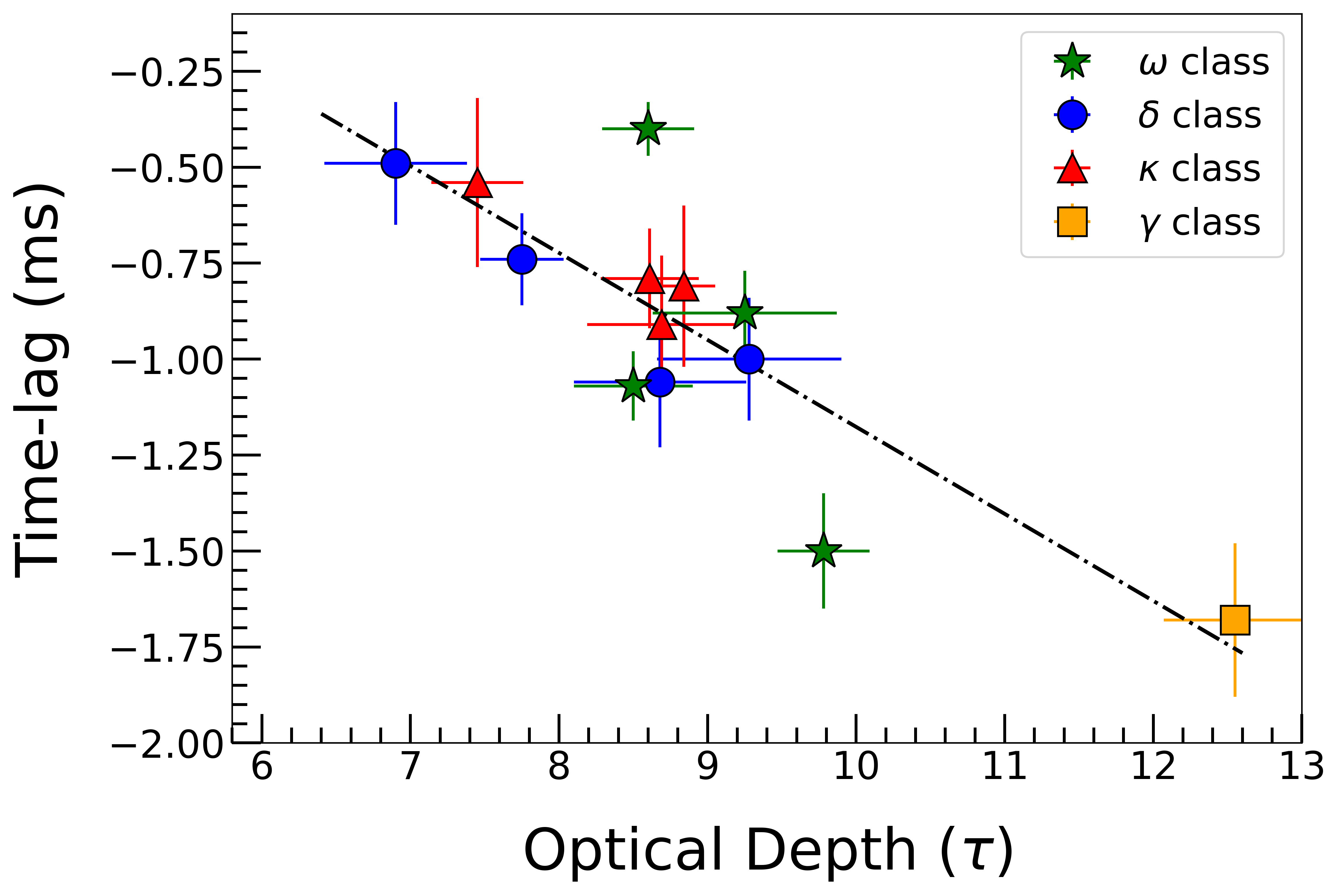}
 \caption{Variation of time-lag is plotted as a function of optical depth ($\tau$). The variability classes are presented in different symbols marked in the inset at the upper right corner. See text for details.}
 \label{fig:lag_tau}
\end{figure}

In this work, we perform a detailed energy dependent variability study for all HFQPOs ($\sim 67$ Hz) observed by \textit{AstroSat}. For the first time, we find soft-lag for all `softer' state variability classes i.e., $\delta$, $\omega$, $\kappa$ and $\gamma$. We also study the correlation of soft-lag with the rms power (Fig. \ref{fig:lag_rms_corr}) and the same with the optical depth (Fig. \ref{fig:lag_tau}) of the Comptonizing region/corona from where the hard photons are being generated. The additional \texttt{powerlaw} could indicate the presence of an extended corona around the central corona \citep{seshadri2022}. It is also evident from the energy dependent power-spectral study that the photons in 6$-$25 keV are responsible for producing HFQPOs \citep{seshadri2022}.

\subsection{The RMS power in HFQPOs }

In order to correlate the time-lag with the rms power of HFQPO, we study the PDS of all observations in the 3$-$60 keV energy band and plotted only for each class in Fig. \ref{fig:pds_plot}. The rms power of the HFQPO lies in the range 0.74$-$2.58\% and is given in Table \ref{tab:pds_table} along with time-lag.
The \texttt{nthComp} flux is maximum in the 6$-$25 keV energy band and the HFQPOs are significant in the 6$-$25 keV energy band, indicating that Comptonized photons are responsible for producing the HFQPOs \citep{seshadri2022}. The rms power increases linearly with energy up to $\sim 20$ keV \citep{seshadri2022} as well observed by \cite{morgan1997} at the first detection of 67 Hz HFQPO using \textit{RXTE} observations. In the \textit{RXTE} era, only 49 observations of GRS 1915+105 exhibit HFQPO at $\sim 67$ Hz out of 1807 observations analyzed by \cite{belloni2013} and they also found that the rms amplitude increases linearly with energy. It is very interesting that none of the HFQPOs were observed in the $\chi$ variability class where a systematic evolution of type C QPOs was observed \citep{dutta2018}. Therefore, the disc flux alone can't explain the observed rms-energy variation though the spectrum belongs to the disc dominated spectral state. 
Thus, it is now obvious that the Comptonization process plays an important role in producing the fast variability in the `softer' classes of the source and HFQPO can be produced due to the oscillation or modulation of the `Comptonizing region’ which may be small and `compact' in size during the disc dominated spectral state \citep{mendez2013,aktar2017,dihingia2019,sreehari2020,seshadri2022}. 
On the contrary, a similar explanation for the case of kHz QPOs in neutron-star systems \citep[see][]{mendez-khzQPO2006,sanna-khzQPO2010} states that if we consider the HFQPOs are produced due to the oscillation mechanism in the disc, the signal/soft-radiation has to be amplified and modulated in the `Comptonizing region’ to contribute hard-lags at the HFQPOs.

\subsection{Time-lag variation with Energy}
 
The energy dependent time-lag study is important to know the exact physical mechanism associated with the generation of HFQPOs. 
In this work, we performed a detailed time-lag study of all HFQPOs observed by \textit{AstroSat} choosing a similar reference band (3$-$6 keV) to compare with \textit{RXTE} observation.
We produce the lag-spectra of 6$-$25 keV photons with respect to the 3$-$6 keV photons because HFQPO can only be seen in this band \citep{seshadri2022}. For the first time, we find soft-lag associated with the HFQPO for all observations during the `softer' variability classes observed with \textit{AstroSat}. This implies that soft disc photons (3$-$6 keV) 
lag behind the harder photons (6$-$25 keV). The hard photons are generally produced via inverse Compton processes \citep{sunyaev_titarchuk1980} and subsequently suffer scattering delay (hard-lag) in respect of soft photons. The soft-lag associated with the HFQPO can be clearly seen in the lag spectra as shown in Fig. \ref{fig:lag-spectra_plot} for all four variability classes.  
The soft-lag for all the classes is found to be in the range 0.40$-$1.68 ms (see Table \ref{tab:pds_table}). Each of the variability class shows a similar lag-energy correlation (see Fig. \ref{fig:lag-energy}) where soft-lag increases with the photon energy implying that higher energy photons always arrive earlier than the lower energy photons. The $\delta$ and $\omega$ classes show maximum soft-lag of $\sim$ 3 ms and $\sim$ 2.5 ms respectively at $\sim$ 18 keV, whereas both the $\kappa$ and $\gamma$ classes exhibit maximum soft-lag of $\sim$ 2.1 ms. 

However, an opposite correlation was reported by \cite{cui1999} and \cite{mendez2013} using \textit{RXTE} observations, i.e., hard-lag at the HFQPO during the similar `softer' variability classes. They found hard-lag (w.r.t 2$-$5 keV) increases gradually with the energy and obtained maximum hard-lag of $\sim$ 5.6 ms and $\sim$ 3.7 ms for $\gamma$ and $\delta$ classes respectively. Thus, a gradual increase of hard-lag with energy, a rise of \texttt{nthComp} flux (see Appendix \ref{section: appendixB}), and an increase in Comptonized photons are observed during the \textit{RXTE} observations.
In contrast, we find an increase in soft photons and increase in soft-lag with energy during the \textit{AstroSat} observations. We also find the presence of reflection features in the \textit{NuSTAR} spectra during the simultaneous observations (see Appendix \ref{section: appendixA}).  
Thus, from the above findings, the observed increase in soft photons during \textit{AstroSat} observations may be attributed to the dominant reflection mechanism, which may lead to the detection of the soft-lag.  This finding contradicts the situation during \textit{RXTE} observations, where a dominant inverse Comptonization mechanism from seed photons in the accretion disc led to the detection of hard-lag. 
However, lack of \textit{AstroSat} observation, poor spectral resolution, and limitations in data products restrict us from probing further to find concrete reasons behind this opposite time-lag feature. Therefore, we need a deeper study of higher spectral resolution (e.g., \textit{NuSTAR}) and extensive analysis of all \textit{RXTE} observations with HFQPOs to find concrete reasons behind this opposite lag behaviour, which is beyond the scope of this paper.

\cite{belloni2019} also found hard-lag for 10$-$20 keV and 20$-$30 keV photons w.r.t 5$-$10 keV photons associated with HFQPO of this source using \textit{AstroSat} observation. 
We also find similar results using the \texttt{LaxpcSoftware} (FORTRAN- based) as obtained by \cite{belloni2019} using the \texttt{GHATS} package, which is IDL-based. Thus, this different lag nature (hard/soft)  observed for the same \textit{AstroSat} observation could be due to the choice of different reference bands.

 A satisfactory explanation \citep{cui1999,nowak1999,poutanen_fabian1999,kara2013,dutta2016,arka-chatterjee2017b,arka-chatterjee2020,nandi_prantik2021} of time-lag includes the major presence of Comptonization, reflection, outflow/jet and the effects of accretion disc geometry.
 However, the hard-lag can be explained by the inverse Comptonization (Compton up-scattering) of soft seed photons from the accretion disc by the energetic electrons in the Comptonizing region or corona \citep{miyamoto1988}. 
 Some of these up-scattered photons or a sudden and small dissipation of hard photons from the corona \citep{malzac2000} may revert to the cooler accretion disc and get reprocessed again. This mechanism is called feedback (or reflection) which could be responsible for producing soft-lag \citep{lee-miller1998}. 
\cite{bellavita2022} showed that the hard/soft-lag depends on the feedback fraction (fraction of up-scattered photons which return towards the disc) which in turn depends on the size of the corona and the inclination of the source. Moreover, the feedback fraction depends mainly on accretion geometry \citep{malzac2000}.
\cite{mendez2013} favoured the Compton up-scattering scenario to explain the soft-lag at 35 Hz and hard-lag at 67 Hz in GRS 1915+105 proposed by \cite{lee_misra_taam2001} where the disc and the corona are coupled and exchange energy.

The observed soft-lag associated with HFQPO can be attributed to the significant dominance of the reflection mechanism during the softer variability classes when corona shrinks (see the lower panel of the diagram in Fig. \ref{fig:Cartoon-diagram}) and the inner edge of the Keplerian disc approaches close to the inner most stable circular orbit (ISCO). The hard photons of higher energy take a longer time to reduce their energy through the down-scattering and hence suffer greater time-lag i.e., soft-lag. This explains the increase of soft-lag with energy as evident from Fig. \ref{fig:lag-energy}.
In the present scenario, the magnitude of the soft-lag (0.40$-$1.68 ms) can be a rough estimator of the distance between the source of photons and the reflector. Here, using time-lag to be light crossing time, we find a wide distance of $\sim 7-27 R_{g}$ (where,  $R_{g}=GM_{BH}/c^{2}$, with $M_{BH}$ referring to the mass of the black hole), which is assumed to produce a significant value of reflection fraction as obtained (see Appendix \ref{section: appendixA}) from the spectral fitting of simultaneous \textit{NuSTAR} observation.

\subsection{Correlation of Soft-lag and HFQPO rms } 

We examine the energy dependent power density spectra and lag spectra of all observations in the 3$-$60 keV energy band to correlate the time-lag with the rms power of HFQPO which is plotted in Fig. \ref{fig:pds_plot}. The rms power of the HFQPO lies in the range 0.74$-$2.58\% and is given in Table \ref{tab:pds_table} along with time-lag.
We, for the first time, find a different linear correlation of time-lag with rms amplitude with slopes $-0.63$, $-0.52$ and $-0.31$ for the $\delta$, $\omega$ and $\kappa$ variability classes respectively, which are shown in Fig. \ref{fig:lag_rms_corr}. However, we detect an outlier among the $\omega$ variability class which was observed on MJD 57995.30 (Orbit 10394). The Pearson correlation coefficients of $\delta$, $\omega$ and $\kappa$ classes (excluding the outlier) are found to be $-0.92$, $-0.97$ and $-0.78$ respectively. The amplitude of soft-lag increases with the increase of rms amplitude of the HFQPO. The higher rms amplitude implies a higher number of differentially intercepted or modulated Comptonized photons, which produces the HFQPOs. The Comptonized photons lead with respect to soft photons that suffer a delay, i.e., soft-lag due to the down-scattering mechanism in the cold accretion disc due to the compact corona and higher inclination of the source. Thus, a larger magnitude of soft-lag would imply a larger delay, which increases with the energy of the photon that suffers more down-scattering.

It is very interesting that an opposite correlation was observed by \cite{zhang2020} for the type-C QPOs in the range of 0.4$-$6.3 Hz where they found that the correlation between QPO fractional rms and the average lag of the QPO can be fitted well with a broken line which implies the amplitude of soft-lag increases with the decrease of rms amplitude and amplitude of hard-lag also increases as rms amplitude decreases.

\subsection{Correlation of Soft-lag and Optical depth}

A wide-band (0.7$-$50 keV) spectral study with thermal Comptonization model \texttt{nthComp} along with a \texttt{powerlaw} component reveals a high value of the optical depth ($\tau$) in the range of $\sim$ 6.90$-$12.55 of the Comptonized medium, close to the black hole. The optical depth can be an important parameter to explain the origin of soft and hard-lag because the opacity of a medium determines the scattering probability as well as the escape probability of the photon. A high value of optical depth ($\tau$) in the range of 6.90$-$12.55 indicates the presence of optically thick medium where the Comptonized photons are generating.
 
A definite linear correlation is observed from Fig. \ref{fig:lag_tau} that the magnitude of the soft-lag increases with the increase in optical depth ($\tau$). The higher value of optical depth ($\tau$) causes a geometrically thick disc and it enhances to occur the Compton down scattering. The higher energetic photons take longer time to lose energy by means of the direct Compton effect. Thus, they suffer a greater delay with respect to higher energy photons before reaching the observer.  
All the variability classes exhibit a similar generic feature. The Pearson correlation coefficient is found to be $-0.83$ which shows that soft-lag and optical depth are strongly correlated. Thus, the spectral study also independently validates the higher value of soft-lag for higher energy photons which we obtained from the temporal study of this source. It is evident from Fig. \ref{fig:lag_tau} that the $\omega$ and $\gamma$ classes exhibit a maximum soft-lag for the highest value of optical depth ($\tau$).

\begin{figure}
 \includegraphics[width=\columnwidth]{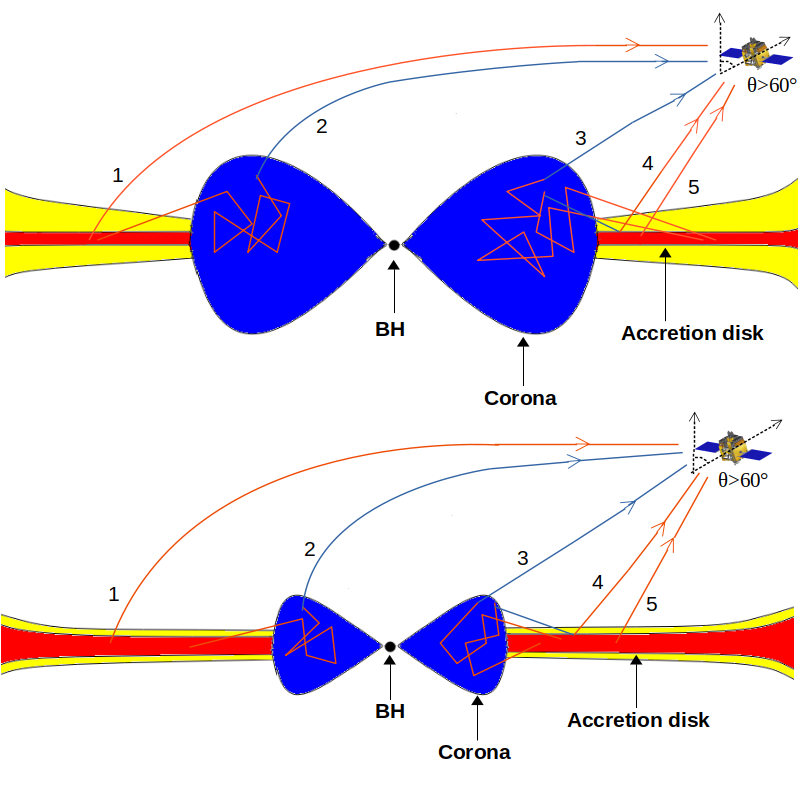}
 \caption{Cartoon diagram of an advective flow disc \citep[red part represents Keplerian flow and yellow part represents Sub-Keplerian flow, see][]{chakrabarti_titarchuk1995} during the `hard' and `soft' spectral state. All possible types of emergent photons from the possible physical processes are considered here. The red and blue curved lines represent the `soft' and `hard' photons respectively. The `1' and `2' lines represent `soft' and `hard' photons coming from the other side of the accretion disc and have suffered gravitational bending. Line `3' goes through inverse Comptonization and reaches the observer directly. Line `4' is the reflected rays from the Keplerian disc. Line `5' represents the soft photons which are generated from the disc and directly reach the observer.
}
 \label{fig:Cartoon-diagram}
\end{figure}

In this present scenario of the `softer' variability classes, the accretion geometry has evolved in such a way that the reflection mechanism dominates over the inverse Comptonization process which can be understood from the cartoon diagram shown in Fig. \ref{fig:Cartoon-diagram}). We expect a large fraction of the photons to come into the Compton down scattering process to produce soft-lag as the feedback fraction increases \citep{bellavita2022} due to the accretion geometry \citep{malzac2000} of the soft spectral states i.e., the smaller size of the corona.

\subsection{Soft-lag and Outflow/Radio emission}
The HFQPOs are only observed from MJD 57503 to MJD 57554 and again from MJD 57943 to MJD 58808 by the \textit{AstroSat} mission till now. The source is found to be in the `softer state' and exhibits four variability classes i.e., $\delta$, $\omega$, $\kappa$ and $\gamma$ during these periods as reported by \cite{seshadri2022}. However, an active radio flux density of $\sim$ 30 mJy was observed only during the period MJD 57600 to MJD 57700 when no HFQPO was noticed
but a tiny radio flux of $\sim$ 1.8 mJy was observed on MJD 58008 \citep[see][for details]{motta2021} when the source exhibits soft-lag at the HFQPO during its $\gamma$ variability class observation.

Generally, the outflow occurs in the hard-intermediate state and due to this outflow, the amplitude of the soft-lag increases \citep{kara2013} as it enhances the number of reflected photons from the Keplerian disc. 
\cite{kim2019} 
showed that most of the out-flowing matter, which has sub-escape velocity, returns to the equatorial region and turns into the return flow. This return flow down-scatters the hard photons to produce soft photons and hence soft-lag. The radio fluxes \citep{vadawale_rao_nandi2001} are associated with the Jet/outflows or return flows \citep{anuj_nandi2001_mnras,anuj_nandi2001_A&A} which contribute to the soft-lag \citep[see][]{patra2019}. In order to explain the soft-lag in the presence of outflow/Jet, we require the detection of any significant radio flux during the HFQPO observation, which is mostly absent in the present cases.

Fig. \ref{fig:Cartoon-diagram} represents a cartoon diagram of the possible accretion scenario in the hard and soft spectral states. The upper figure represents accretion geometry during the hard state when gravitational bending effects \citep[see][]{arka-chatterjee2017a} and reflection mechanism \citep{poutanen_fabian1999,dutta2016} are not dominating as the softer spectral state (lower figure). 
The possible effects mentioned above are only dominating for the edge-on (i.e., high inclination angle) sources in soft state. 
\cite{arka-chatterjee2017a} found that gravitational bending enhances the number of hard photons with inclination angle, whereas the multicolor blackbody part of a spectrum becomes flat at higher inclination angles. It is a clear effect of the focusing due to photon bending. In this scenario, the inversely Comptonized photons (i.e., the photons that originated in the Keplerian disk are intercepted by the corona) naturally lag behind soft photons that are coming directly to the observer, and this mechanism dominates with the larger size of the Comptonizing region/corona (during hard spectral state). However, the Compton down-scatter mechanism dominates with the source's increasing inclination and decreasing corona size. This direct Compton effect will eventually enhance the reflection mechanism over all other physical mechanisms to produce soft-lag.

\section{Conclusions} \label{section: conclusions}

The time-lag property in the wide band energy range (3$-$60 keV) is important to understand the accretion geometry and to identify the dominant physical processes that produce HFQPOs. GRS 1915+105 exhibits a generic HFQPO $\sim$ 67 Hz persistently for over the last 25 years. However, a fewer observations during the `softer' variability class have been detected the HFQPO using \textit{RXTE} and \textit{AstroSat} observation. Previously, only hard-lag was detected using the \textit{RXTE} observation but we discovered soft-lag associated with HFQPOs.
Based on our findings, it can be inferred that the observed soft-lag during \textit{AstroSat} observations can be attributed to the dominant reflection mechanism, whereas a dominant inverse-Comptonization mechanism led to the detection of hard-lag in the similar `softer' variability class during \textit{RXTE} observations. We summarised here a generic feature of time-lag associated with the HFQPOs during four variability classes i.e., $\delta$, $\omega$, $\kappa$ and $\gamma$ using all \textit{AstroSat} observations.

\begin{itemize}
\item The first detection of soft-lag ($\sim$ 0.40$-$1.68 ms) of the 6$-$25 keV energy band with respect to 3$-$6 energy keV associated with HFQPOs. All four `softer' state variability classes exhibit a coherent lag-energy correlation. 

\item The energy-dependent time-lag study shows the increase in soft-lag with energy for the first time and exhibits maximum soft-lag of $\sim$ 3 ms and $\sim$ 2.5 ms for the $\delta$ and $\omega$ class, whereas the $\kappa$ and $\gamma$ both classes show a maximum soft-lag of $\sim$ 2.1 ms.

\item We, for the first time, find a correlation of time-lag with rms amplitude for each of the variability class i.e., $\delta$, $\omega$ and $\kappa$ where the amplitude of soft-lag increases with the rms amplitude of the HFQPOs. 

\item Spectral study suggests a high value of the optical depth ($\tau$ $\sim$ 6.90$-$12.55) of the Comptonized medium close to the black hole. A linear correlation is observed for all variability classes where the magnitude of the soft-lag increases with the increase in optical depth ($\tau$). 

\item No significant radio-emission was observed when HFQPOs are detected. However, a tiny radio flux of $\sim$ 1.8 mJy was observed during the $\gamma$ variability class.
  
 \end{itemize}

\section*{Acknowledgments}
Authors thank the anonymous reviewer for valuable suggestions and comments that helped to improve the quality of this manuscript. BGD, PM and AN acknowledge the support from ISRO sponsored project (DS-2B-1313(2)/6/2020-Sec.2). PM, BGD thanks the Department of Physics, Rishi Bankim Chandra College for providing the facilities to support this work.  BGD acknowledges `TARE’ scheme (Ref. No. TAR/2020/000141) under SERB, DST, Govt. of India and also acknowledges Inter-University Centre for Astronomy and Astrophysics (IUCAA) for the Visiting Associate-ship Programme. AN thanks GH, SAG; DD, PDMSA, and Director, URSC for encouragement and continuous support to carry out this research. This work uses the data of \textit{AstroSat} mission of ISRO which is archived at the Indian Space Science Data Centre (ISSDC). We have used the data of Soft X-ray Telescope (SXT) which is calibrated and verified by \textit{AstroSat-SXT} team. We thank the SXT-POC team at TIFR for providing the necessary software tool to analyse \textit{SXT} data. This work has also used \textit{LAXPC} data which is verified by LAXPC-POC at TIFR. We thank \textit{AstroSat} Science Support Cell for providing the software \texttt{LAXPCsoftware} for the analysis of \textit{LAXPC} data.

\section*{Data Availability}
Observational data used for this publication are available at the Astrobrowse (AstroSat archive) website \\ 
\url{https://webapps.issdc.gov.in/astro_archive/archive} of the Indian Space Science Data Centre (ISSDC).



\bibliographystyle{mnras}
\bibliography{reference}


\appendix

\section{Reflection feature in the softer variability state with \textit{NuSTAR} spectrum} 
\label{section: appendixA}

\begin{figure}
 \includegraphics[width=\columnwidth]{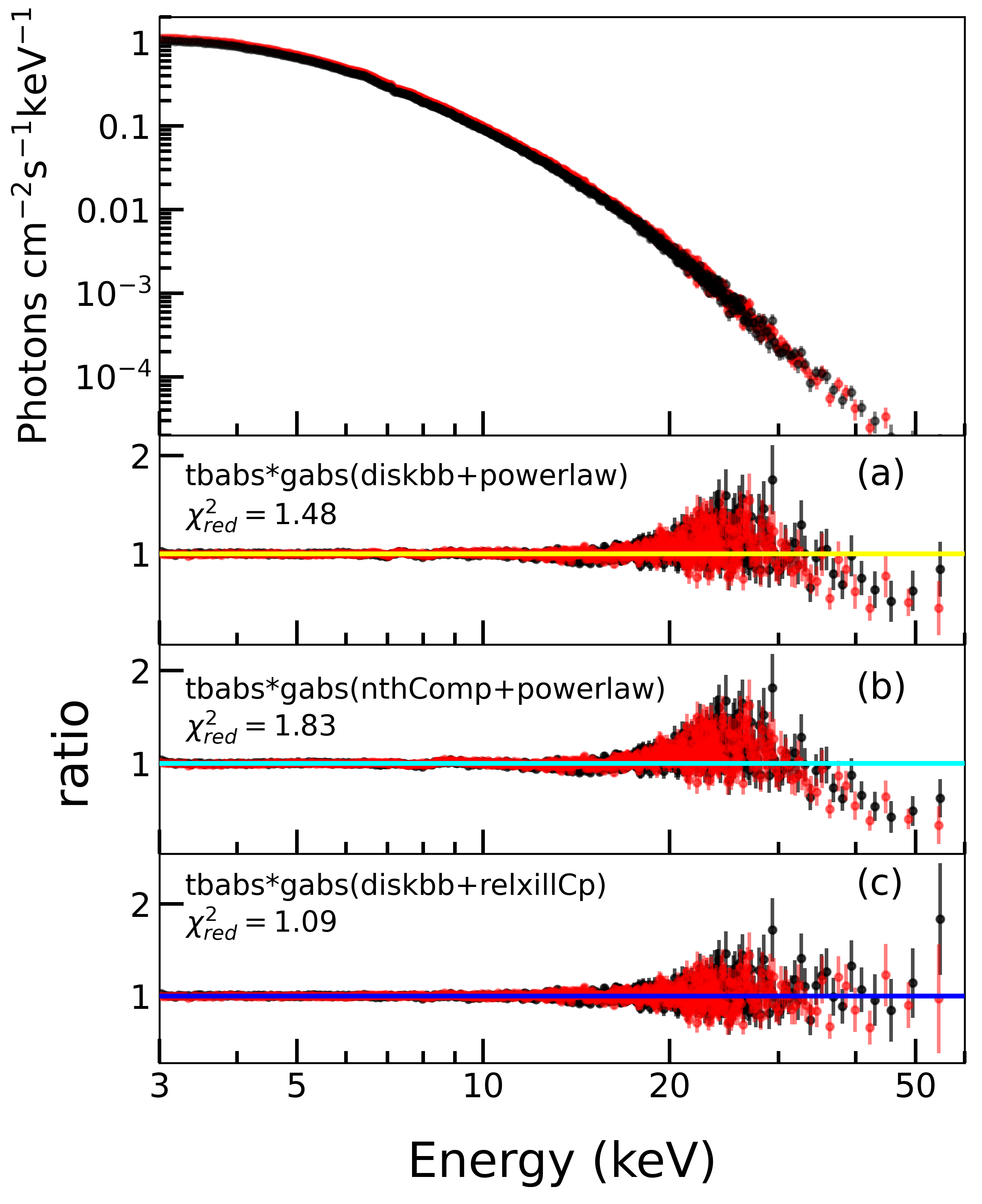}
 \caption{FPMA (black) and FPMB (red) energy spectra observed on MJD 57553 and the ratio of the model M1, M2 and M3 shown in panel (a), (b) and (c) respectively. A reflection signature is visible in (a) and (b) at the energy range 20 – 30 keV. The model combination and the $\chi^{2}_{red}$ are mentioned in the upper left corner of each panel.}
 \label{fig:spectrum_nustar}
\end{figure}

We have considered a simultaneous \textit{NuSTAR} observation (Obs ID 30202033002) with one of the analyzed observations of \textit{AstroSat} (Orbit 3860, MJD 57553, $\delta$ class) to find the reflection features in the spectrum. We have tried to get the Compton hump at high energy in the spectrum of \textit{LAXPC20}  using \texttt{Tbabs $\times$(diskbb$+$powerlaw)} model, but unable to detect any hump-like feature in the energy range 20 – 30 keV due to the poor spectral resolution of \textit{LAXPC}. We have extracted the cleaned events using \texttt{nupipeline} and obtained both \textit{FPMA} and \textit{FPMB} spectra using \texttt{nuproducts}. We fitted the \textit{NuSTAR} spectra considering the model \texttt{TBabs$\times$gabs(diskbb$+$powerlaw)} (referred to as M1 hereafter). A sharp absorption feature is observed in the residual at $\sim$ 7 keV, which is taken care of using the Gaussian absorption model \texttt{gabs}. The best-fit yields a $\chi^{2}_{red}$ of 1.48. We observed a Compton hump in the ratio plot in the energy range 20$-$30 keV (see panel (a) in Fig. \ref{fig:spectrum_nustar}). This signature implies that the reflection feature is present in the spectra. 
We consider another model \texttt{TBabs$\times$gabs(nthComp$+$powerlaw)} (M2 hereafter), to incorporate the reflection feature in the spectrum. In this case, the best-fit yields a $\chi^{2}_{red}$ of 1.83. The Compton hump in 20 – 30 keV is also observed in this model in the ratio plot (see panel (b) in Fig. \ref{fig:spectrum_nustar}).

Further, we use a self-consistent reflection model \texttt{relxillCp} \citep{dauser2022} which can take care of both Comptonised emission from corona and the reprocessed emission from the disk. The reflection model \texttt{relxillCp} is considered to study the reflection feature in the spectrum quantitatively. This model assumes the primary source spectrum model to be \texttt{nthComp}. We fit the wide-band energy spectra considering the model \texttt{TBabs$\times$gabs(diskbb$+$relxillCp)} (M3 hereafter), and we obtained a satisfactory fit (see panel (c) in Fig. \ref{fig:spectrum_nustar}), which yielded a $\chi^{2}_{red}$ to 1.09. The inclination of the source is frozen to the standard value of 65$^{\circ}$. The inner disk radius is assumed to be three times $R_{ISCO}$ of the black hole according to the norm of \texttt{diskbb} obtained from the fitting in model M1. The spectral fit suggests that the accretion disk is highly ionized. The other parameters are considered according to \cite{bhuvana2023_relxill}. The model parameter, reflection fraction, is found to be 0.48.

The reflection fraction is defined as the ratio of the intensity of the primary source (\texttt{nthComp} flux due to the spherical corona) irradiating the disk and the intensity directly going to infinity \citep{dauser2016}. The non-zero value of the reflection fraction shows the presence of reflection features in the spectrum. Thus, the existence of the reflection mechanism can be concluded during the soft-state of the accretion disk dynamics.

\section{Evolution of energy spectra of \textit{RXTE} and \textit{AstroSat} observations}
\label{section: appendixB}
We considered five observations for spectral analysis, two of which are from \textit{RXTE} \citep[]{cui1999, mendez2013} where hard-lag (w.r.t 2$-$5 keV) was observed at the HFQPO. The rest of the observations are from \textit{AstroSat} where soft-lag is detected (w.r.t 3$-$6 keV). We considered the same spectral model prescription, 
\texttt{constant$\times$TBabs$\times$(nthComp + powerlaw)}
as mentioned in \S \ref{subsec:spectral_analysis}, and the fitted parameters are shown in Table \ref{tab:spectral_parameter2}. The model parameter $kT_{bb}$ was kept constant at 0.1 keV for all \textit{AstroSat} observation, as \cite{sreehari2020} found the $kT_{bb}$ to be in the range of 0.1$-$0.3 keV. However, we could not constrain the $kT_{bb}$ parameter in the case of \textit{RXTE} observations. Further, we calculated the flux contribution in the energy range of 3 – 50 keV due to \texttt{nthComp} and \texttt{powerlaw} separately.

\begin{table*}
	\centering
	 \caption{\label{tab:spectral_parameter2}Model parameters of the \textit{RXTE} and \textit{AstroSat} observations fitted with model \texttt{constant$\times$TBabs$\times$(smedge$\times$nthComp + powerlaw)}}.  
	 \resizebox{19cm}{!}{
	 \begin{tabular}{cccccccccccc}
	 \hline
	 \hline

Mission & Obs ID/Orbit & MJD & $kT_{e}$ & $\Gamma_{nth}$ & $kT_{bb}$ & $\Gamma_{PL}$ & $\chi^{2}_{red}$ & $Flux_{nth}$ & $Flux_{PL}$ & $Flux_{nth}\%$ & Class\\
  &   &   & (keV) &   & (keV) &   &   & ($\times 10^{-8} $erg$ cm^{-2} s^{-1}$) & ($\times 10^{-8} $erg$ cm^{-2} s^{-1}$)\\
    \hline
    \hline
\textit{RXTE} & 10408-01-06-00 & 50208.58 & 3.54$\pm$0.21 & 3.10$\pm$0.10 &  1.33$\pm$0.01 & 3.00* & 0.98   & 3.96   & 0.02  &  99.49  & $\gamma$   \\

\textit{RXTE} & 80701-01-28-01 & 52933.69 & 2.95$\pm$0.10 & 2.28$\pm$0.07 & 1.12$\pm$0.06 & 2.63$\pm$0.04 & 0.90  & 4.08 & 0.25 & 94.22 & $\delta$   \\

\textit{AstroSat} & 3860 & 57553.88 &  2.83$\pm$0.06 & 2.45±0.03  & 0.1* & 3.19$\pm$0.01 & 1.11 &  2.31  & 0.96  &  70.64 & $\delta$               \\

\textit{AstroSat} & 9895 & 57961.42 & 2.83$\pm$0.11 & 2.27$\pm$0.06 & 0.1* & 3.07$\pm$0.03 & 1.09 & 1.20 & 0.48 & 71.42 & $\kappa$        \\

\textit{AstroSat} & 10583 & 58008.08 & 2.40* & 1.85$\pm$0.04 & 0.1* & 2.90$\pm$0.01 & 0.95 & 2.24 & 0.96 & 70.00 & $\gamma$   \\

    \hline
 \end{tabular}
}	 
	 \begin{list}{}{}
		\item[*]Parameter is fixed.
	\end{list}
\end{table*}

We find that the flux contribution of \texttt{nthComp} in the \textit{AstroSat} observation has decreased nearly 30\% from the \texttt{RXTE} observation. The reduction of the \texttt{nthComp} flux that we observed during the \textit{AstroSat} observation implies a smaller number of Comptonised photons to the observer. Here, we expect a fraction of the Comptonised photon to get down-scattered by the Keplerian disk before coming to the observer and subsequently suffer a delay (soft lag) with respect to the Comptonised photons that come directly.


\bsp	
\label{lastpage}
\end{document}